# Towards integrated metatronics: a holistic approach on precise optical and electrical properties of Indium Tin Oxide

*Yaliang Gui[1], Mario Miscuglio[1], Zhizhen Ma[1], Mohammad T. Tahersima[1], Shuai Sun[1], Rubab Amin[1], Hamed Dalir[2], and Volker J. Sorger[1,*]*

[1]Department of Electrical and Computer Engineering, George Washington University, Washington, DC 20052, USA
[2]Omega Optics, Inc. 8500 Shoal Creek Blvd., Bldg. 4, Suite 200, Austin, Texas 78757, USA
*sorger@gwu.edu

**ABSTRACT**

**The class of transparent conductive oxides includes the material indium tin oxide (ITO) and has become a widely used material of modern every-day life such as in touch screens of smart phones and watches, but also used as an optically transparent low electrically-resistive contract in the photovoltaics industry. More recently ITO has shown epsilon-near-zero (ENZ) behavior in the telecommunication frequency band enabling both strong index modulation and other optically-exotic applications such as metatronics. However the ability to precisely obtain targeted electrical and optical material properties in ITO is still challenging due to complex intrinsic effects in ITO and as such no integrated metatronic platform has been demonstrated to-date. Here we deliver an extensive and accurate description process parameters of RF-sputtering, showing a holistic control of the quality of ITO thin films in the visible and particularly near-infrared spectral region. We further are able to custom-engineer the ENZ point across the telecommunication band by explicitly controlling the sputtering process conditions. Exploiting this control we design a functional sub-wavelength-scale filter based on lumped circuit-elements, towards the realization of integrated metatronic devices and circuits.**

Introduction

Indium Tin Oxide films have been extensively employed for diverse applications in the fields of optics and electronics in both research and industry [1]. Multiple uses of ITO ranges from photovoltaics [2–5] to conductive displays [6], integrated photonics [7, 8] and solid-state [9]. ITO's versatility can be attributed to its concurrent optical transparency (T) and sheet resistance ($R_{sheet}$) (Figure of merit = T/ $R_{sheet}$)[10], as well as compatibility with the established technology [11]. Moreover, ITO's carrier density can be electrostatically tuned similarly to metal-oxide-semiconductor (MOS) capacitors, simply by applying a bias voltage. This carrier modulation translates into either a refractive index or absorption change [12]. For this reason, ITO has been largely employed as an active material in electro-optics modulator [13–20] in integrated photonics devices and plasmonic metasurfaces [21–24], specifically displaying high performance optical electro-refractive modulation [25]. In this regard, recently, new possibilities are emerging to use ITO as an active material in sub-wavelength waveguide integrated electro-optics modulators, which showed high performances regarding extinction ratio dB/μm [7, 26–28]. Nevertheless, according to the application, the fundamental challenge in processing ITO is concurrently obtaining thin films with favorable electrical and optical conditions. For instance, in order to design efficient, transparent conductors for photovoltaic and conductive display applications, relatively low resistivity and optical transparency have to be achieved. For electro-optics application, the wavelength of the epsilon-near-zero (ENZ), the carrier concentration and absorption losses, which define the modulation strength, need to be accounted. Moreover, ITO films could enable an ENZ circuit board and nanoscale structured ITO RCL-equivalent circuit elements, thus creating a viable means to realize metatronic circuits [29]. Recently [30], ITO films have exhibited an extremely large ultrafast third-order nonlinearity at ENZ wavelengths, which could be exploited in nonlinear nano-optics application. Therefore, control over ENZ wavelength could constitute a fundamental building block towards the realization of metatronic devices [31] as well as nonlinear optical components. However, ITO's material parameters such as resistivity and permittivity are complex depending on a multitude of process conditions. For this reason, obtaining a thorough understanding of the effects of specific process parameters to the electro-optical properties of the ITO films is highly desirable and still represents an open challenge. Recently system parameters on different deposition techniques such as PEMOCVD [32], IBAD [33] and DC sputtering [34, 35] have been developed in an ad-hoc manner, however a significantly cross-validation process for the RF sputtering process and subsequent thermal treatment is still lacking. While previous work investigated selected material properties, a holistic understanding of a) the inter relationship among these material parameters and b) the precise process control thereof is far from being completely understood. In addition a viable path to realize metatronic circuits offering the best of both photonic signal communication and nanoscale component density is yet outstanding. Here we show the holistic control of the quality of ITO thin films in the visible and particularly near-infrared spectral region. We alter key material-film property factors during a RF-sputtering processing parameters such as the oxygen concentration and thermal annealing. Using this approach, we thence develop an interdependently-complete set of metrology processes for determining the conductivity, carrier density, and mobility of these ITO films and bring them in congruence with values obtained from spectroscopic ellipsometry. We discover a cross-sectional film thickness-dependent carrier distribution as a function of thermal carrier activation, and demonstrate ability to tune the resistivity gradient by restoring the film crystalline structure upon thermal treatment. We further demonstrate the precise control over ENZ film properties across the telecommunication band, which enables engineering a variety of optoelectronic devices with exotic properties. Using this process control in ITO, we analyze a functional sub-wavelength-scale filter based on lumped circuit-elements in a metatronic approach. Such self-consistent process and precise control of ITO's properties electrical, spectral, and optical parameters serves a wide application space such as in optoelectronics, metamaterials, metatronics, quantum technologies, energy-harvesting, and sensing.

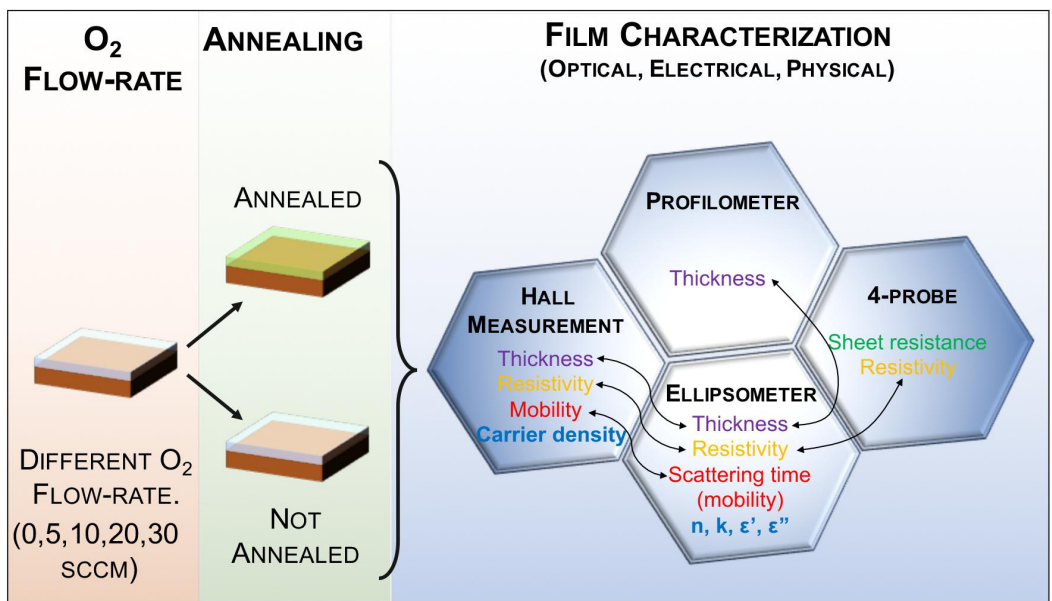

Figure 1. Flow chart of the ITO film process and holistic characterization. Different films are sputtered at different Oxygen flow-rate conditions (0, 5, 10, 20, 30 sccm). Post annealing process in a sealed chamber in inert atmosphere ($H_2$ and $N_2$). Thickness is directly measured through profilometer. Validation of the electro-optical properties of ITO films through complimentary set-ups which includes Hall, 4-Probe and Transmission line measurements. The black arrows illustrate the cross-validation mechanisms opted in this study.

**ITO deposition and characterization: Validation process**

We initially focus on defining a consistent and tunable ITO deposition process using RF sputtering, which allow to finely adjust the electrical and optical properties of the film. The purpose of our study is ultimately to engineer ITO film properties in order to deliver well-defined control over critical parameters for obtaining films that could enhance fabricated device properties.

In our experiments, films are deposited on 1x1 cm$^2$ Si/SiO$_2$ substrates, utilizing reactive RF sputtering (Denton Vacuum Discovery 550 Sputtering System). ITO sputtering targets consist of 10% SnO$_2$ to 90% In$_2$O$_3$ by weight (refer to the method section for details). Several replicated samples are produced for each oxygen flow-rate condition (0, 5, 10, 20, 30 sccm) and sputtering time (500,1000,1500,2000 s), for 40 sccm Argon flow-rate (50 sccm Ar flow rate study is presented in SI). For each condition, represented by a single data-point, the study is repeated four times for statistical purpose and repeatability. Conducive to promote crystallinity and study the charge carrier activation mechanisms, a sub-group of deposited films is subjected to a thermal treatment at 350$^o$C in the inert atmosphere (a mixture of H$_2$ and N$_2$) for 15 minutes. Contrary to atmosphere annealing treatment [46], which aims to promote optical transmittance by filling vacancies, an annealing treatment in inert conditions improves both the film conductivity [47,48]. Differently from other studies [32-35], hereby we map the overall conductivity and optical absorption variation for different oxygen flow rates and deposition time. After deposition, the step height of ITO films is measured in multiple areas for estimating the uniformity of the deposition and determining the initial fitting parameters in our spectroscopic ellipsometry.

For assessing the quality of our process and determining its reliability, direct and indirect measurements are used as rigorous cross validation tools. The resistivity of the thin film is indirectly obtained via ellipsomety and, directly, through 4-probe station (Four Dimensions 280DI) and Hall effect measurement system (HMS-5500). Hall measurement also gives information on the carrier type (n-doped) and concentration as well as the mobility.

A straightforward Drude model, as well as the material model provided by Genosc library, is reductive and inaccurate, leaving aside a complete optical description of the film in the wide spectral window considered, under these process conditions. Relating to this, we use a fitting sequence which comprehends Cauchy, B-spline, Drude and Lorentz models, [36] allowing inter alia, an accurate description of the electrooptical features of the film as function of its depth (see supplementary information). This rigorous model for fitting the optical spectra enables the description of the refractive index and optical conductivity, as well as mobility and carrier concentration, which is cross validated by hall bar and 4-probe direct in-situ measurements (Fig. 1). The epsilon near-zero position is evaluated from the complex permittivity by ellipsometry.

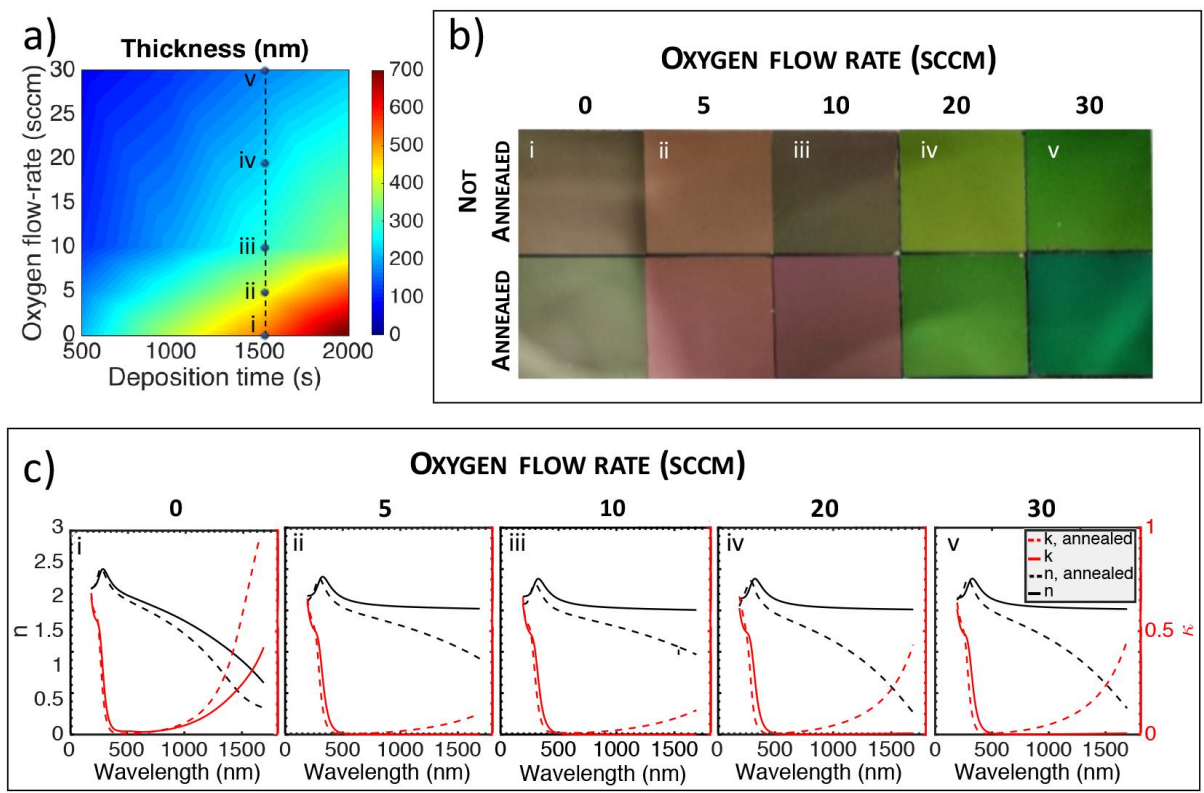

Figure 2. Oxygen flow-rate and post deposition heating treatment effects: ITO film Thickness as function of the Oxygen flow-rate (0,5,10,20,30) for different deposition time (500s, 1000s, 1500s, 2000s). Image of the 1x1 cm$^2$ ITO films on SiO$_2$/Si substrate for different oxygen flow-rate, before and after annealing. Spectral behavior of the real (black) and imaginary (red) part of the refractive index of films sputtered using different oxygen flow-rate and effects of the post-deposition heat treatment process (dashed lines).

**Effects of the Oxygen flow rate and thermal treatment on dispersion and absorption**

In order to establish baselines control over ITO films, we aim to understand the thickness-related influence on material parameters. We find that thinner films are obtained for higher oxygen concentration as a result of reduced sputter material deposition rate (Fig. 2a). Quantitatively we find that for a fixed deposition time (e.g. 1500s shown in S1 in Supplementary information) the ITO film thickness decreases by: $t=ae^{bx}$ being t the thickness, a=509nm the y-intercept and b=0.047sccm 1 the decay constant (RMS 97%). The deposition conditions, summarized in the method section, enable a precise control of the thickness, with a repeatability within 10 nm (see SI) for the thickest film deposition (0 sccm, 2000 s). The films show a maximum roughness of around 3 nm. Taking all measured thicknesses measured using ellipsometry and profilometer, we find a consistent level within a 95% agreement. No substantial variation of the thickness is observed for thermally treated and untreated samples (SI Fig S1). In the interest of simplicity, we focus our discussion for the remainder of this paper on films evaporated for 1500 s, nevertheless, our findings are equally valid for other deposition times, which are summarized in the supplementary information, where an overall summary is provided. The stark contrast in obtainable ITO material properties discussed further below can even be seen visually (Fig.2b) and hints towards a strong ability to engineer the material properties. Four replicated samples are deposited for each experimental condition (i.e. Oxygen flow-rate, deposition time). The repeatability of the process is ensured by the concurrent low variability of the thickness and optical constants (well below 5%). A qualitative analysis on the film colorizations actually provides first insights on the doping type of the material [50]; a tendency to a brown color (Fig2b i,ii) indicates a higher Sn-In doping while a green-yellow colorization corresponds to the presence of oxygen vacancies (Fig2b iv,v). However, in order to quantitative analyze the optical constants, spectroscopic ellipsometry studies are reported in Fig 2c i-v for 1500 s deposition time. Beside 0 sccm, it appears that ITO films not subjected to an annealing process, films are predominantly not absorptive above 500 nm. In the wavelength range from 500 nm to 1680 nm, values of extinction coefficient are, in fact, below hundredths. A major effect attributed to thermal annealing is in fact the activation of the carrier [35], which is responsible of the optical absorption. As a consequence of that, films with 0 sccm (Fig 2c i) appear to be strongly absorptive in this region, due to the higher oxygen vacancies and consequent higher carrier concentration. Films deposited with this oxygen flow-rate could be embedded in NIR metamaterial perfect absorbers [51] or as a building block in meatasurface [52]. Consistently, for all the different oxygen flow-rates tested, after annealing, ITO thin films appear to be more absorptive. The value of $\kappa$, the imaginary part of the refractive index, of annealed samples substantially increases as a function of the wavelength in the IR region, compared to not annealed samples, thus producing a shift in the plasma frequency and an overall variation of the refractive index, being $n(\omega)$ and $\kappa(\omega)$ in Kramer-Kronig relation. The filling of the oxygen vacancies at 5 and 10 sccm (Fig 2 c iii-iv) induces the sputtered ITO films to show the lowest $\kappa$ among the other groups and they might be a viable option for low-loss conductive transport layer in optoelectronic devices or as a sensing platform [49]. For higher oxygen flow rate (20, 30 sccm) the films become more absorptive in the IR.

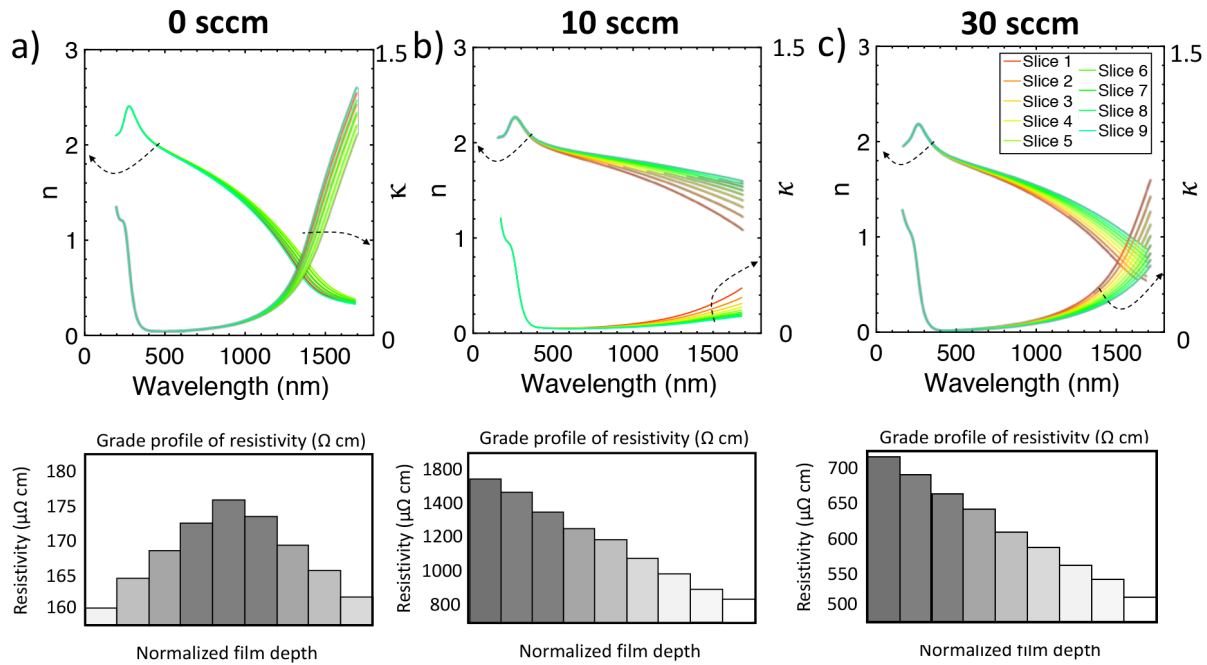

Figure 3. Complex refractive index and resisitivity depth profile a-c) Real (left y-axis) and imaginary part (right y-axis) of the refractive index spectral response investigated by ellipsometry. Different curves (red to green) represent different depth profile distribution within the film thickness (red/top layer, green/bottom layer) for 0 (a) , 10 (b), 30 (c) sccm after annealing. Bottom row represents the resistivity depth profile along the film thickness.

Post-deposition annealing profoundly changes the structural and optical properties of ITO thin films, as previously discussed in the optical characterization. This process, in fact, represents an effective way to promote crystallinity and modify the physical features of ITO films, such as roughness, but also it contributes in drastically altering the carrier distribution. Our ellipsometry studies uses a fitting approach which contemporary minimize the error (RMS) computed in the fitting and takes into account a graded variation of optical/electrical parameters. This approach is essential for quantitatively pointing out that the resistivity, for annealed samples, is a function of the film depth as discussed below, which confirms that the thermal processes can favor the redistribution of carriers, homogenizing them throughout the sample, as illustrated by Buchanan et al [37]. Another effect, which can be ascribed to thermal heating, is related to optical absorption. Moreover, for oxygen flow-rates above 0 sccm, the absorption significantly increases in the IR region. As an obvious influence on the resistivity, this type of films displays an exponential decaying resistivity within the film depth (Fig. 3 b-c). Interestingly, the absorption and resistivity trend for film sputtered with a null Oxygen flow-rate (Fig. 3 a) have the highest value confined in the film core and it is significantly smaller (1 order of magnitude) than highest resistivity recorded for the ITO film sputtered at 10 sccm oxygen flow-rate [44]. We can speculate that being the thickest (approx. 600 nm) among the studied groups, the annealing time is not sufficient to completely redistribute the carriers throughout its depth.

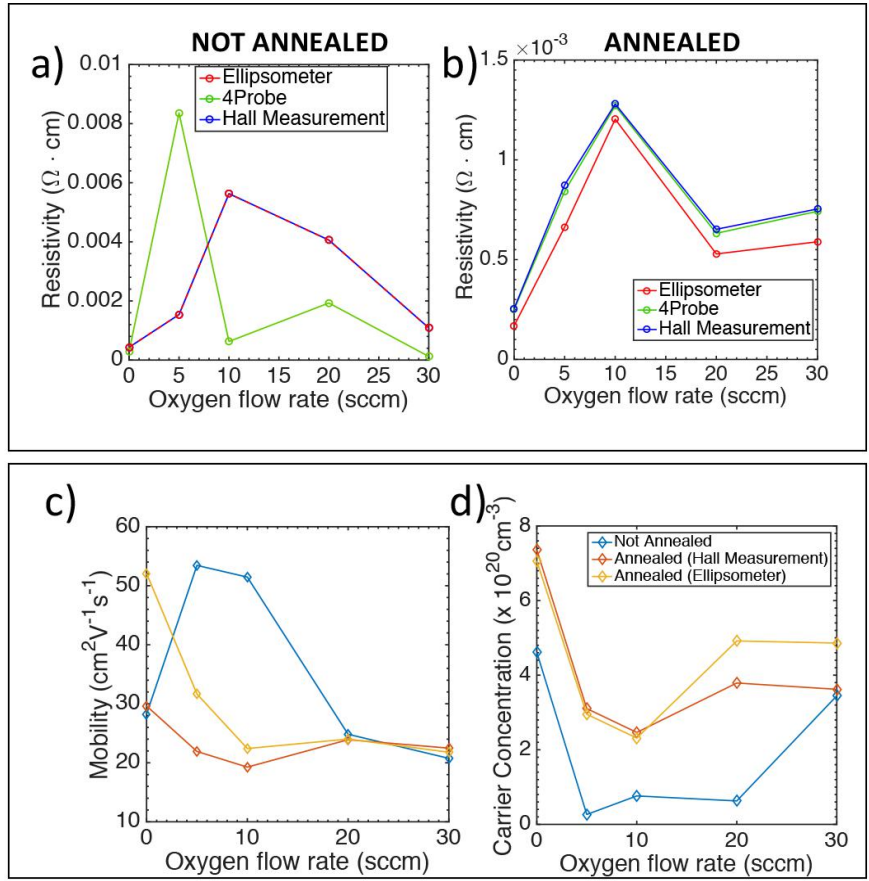

Figure 4. Resistivity measured with different techniques (Ellipsometer, 4-Probe and Hall Measurement) as function of the Oxygen flow-rate thermally untreated (a) and treated (b) samples. Mobility and carrier concentration for treated and untreated samples validated through direct (Hall Effect) and indirect (Ellipsometer) measurement

**Effects of the Oxygen flow rate and thermal treatment on the electrical properties**

Electrical properties, carrier concentration and mobility of ITO have been found previously to vary for different deposition techniques, but also for different process conditions for the deposition such as power, oxygen flow, and annealing temperature [3, 35, 38, 39]. In this section we focus on the impact of the oxygen flow-rate and thermal post deposition treatment on the electrical properties of ITO. Figure 4 summarizes the electrical properties of the ITO films sputtered with different oxygen flow rate (1500s), for thermally untreated (a) and treated samples (b), measured using three different methods, i.e. Hall measurement, 4-probe and spectroscopic ellipsometry. It is worth noticing that the measurements (direct and indirect) are in quasi-perfect agreement (>90% correlation). Only in case of the 4-probe measurement of non-annealed samples, the resistivity has a different trend. This has to be attributed to the instability of the contacts-film junction, due to the rather soft upper layer of the samples, which are not thermally treated. It is evident that not annealed thin films are less conductive, displaying one order of magnitude higher resistivity (a-b), an overall slightly higher mobility (c), lower carrier concentration (d) compared to the annealed samples. Therefore annealed films are generally more suitable for the implementation of capacitive sensors, and promotes the absorption in the optical telecom wavelength, enabling the fabrication of efficient electro-absorption modulators based on ITO films. As previously mentioned, for ITO, electrons are the majority carriers and they are originating mainly from the doping donor Sn and oxygen vacancies. For our sputtering conditions we show that increasing oxygen flow-rate produces an initial increase of the resistivity, induced by a lower carrier concentration. The lower carrier concentration is reached for Oxygen flow rate within 5-20 sccm. Therefore, increasing the oxygen flow rate replenishes the oxygen vacancies up to 10 sccm. For higher oxygen flow-rates the sputtered particles from the target cannot oxidize sufficiently, hence the ITO films are anoxic and sub-oxides such as $InO_x$ and $SnO_x$ are present in the films [40].

For annealed sample both the carrier concentrations and mobility overall decrease by increasing oxygen flow rates [41]. This can be attributed to the concurrent filling of the oxygen vacancies and the deactivation of the Sn donor by the overflowing oxygen. It can be said that 10 sccm of oxygen flow rate can be an optimum flow rate for obtaining high resistivity (low carrier concentration) and as previously shown low absorption.

The mobility on the effect of annealing is within the same order of magnitude, although the trend is rather different. The mobility in non-annealed films is higher for lower carrier concentrations, while opposite trend is visible for annealed films.

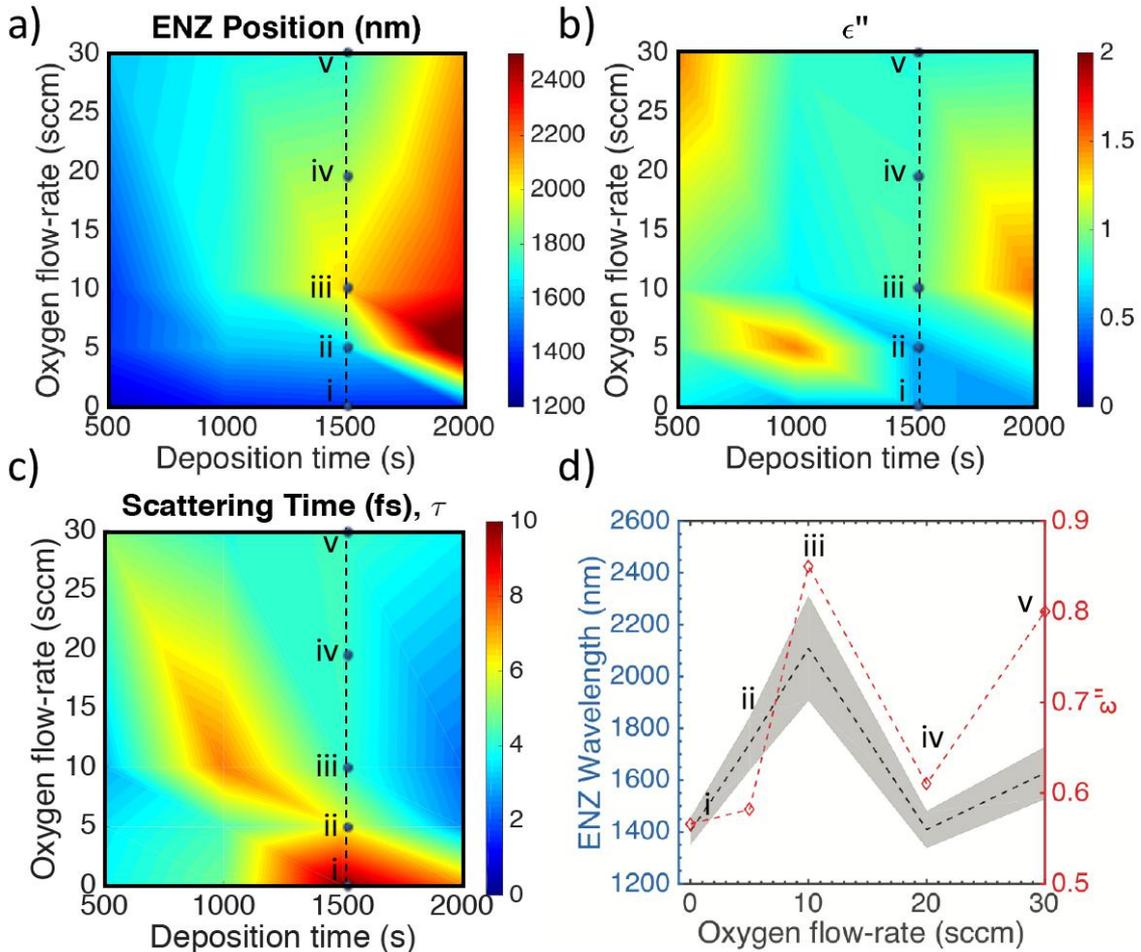

Figure 5. Experimental (interpolated) data of ITO film loss and ENZ-position as a function of oxygen process flow-rate. a) Measured ENZ wavelength as function of the deposition time and oxygen flow-rate b) Measured damping ($\varepsilon''$) function of the deposition time and oxygen flow-rate c) Scattering Time t as function of the deposition time and oxygen flow-rate. d) ENZ wavelength (black solid line, left-axis) as function of the Oxygen flow-rate and corresponding damping ($\varepsilon''$)) (red solid line, right-axis) for ITO film deposited for 1500 s

Afterwards, for annealed samples, we investigate the variation of spectral response of the real part of the permittivity (e') for different oxygen flow-rate and deposition time. The x-intercept of the curve reveals the wavelength where $\varepsilon' = 0$, i.e. ENZ (see supplementary information). The trend of ENZ position as function of the oxygen flow-rate is strongly inherited from the resistivity response (Fig. 5 b), displaying a 96% cross correlation. This is further illustrated by the difference scattering time t which has complementary trend with respect to the $\varepsilon''$ as a function of deposition time and oxygen flow-rate. Our ability, to methodically tune ENZ position for a broad range of wavelength spacing from the E- to L-bands (1400 to 2500 nm) offers to engineer precise ENZ-based devices at targeted wavelengths (Fig.5 d). Considering 1500s deposition time, the longest wavelength reached is for 10 sccm, which corresponds to lower absorption, as well as higher resistivity. As a matter of fact, this further degree of freedom in setting the ENZ position could for instance enable the realization of sub-wavelength electooptic modulators based on ITO films embedded in integrated photonics circuit [27]. Through this study, RF-sputtered ITO films with crafty ENZ positioning in the optical telecom range can be deposited, thus lowering the energy required to switch from ENZ (high absorption) and epsilon-far-from-zero (low absorption) [45]. An intriguing field that could greatly benefit of this study is metatronics, allowing the implementation of nano-optical circuit entirely based on ITO film, wisely doped, using suitable process parameters [42]. For illustrating the potentiality of this study, we hereby demonstrate the possibility to finely tune RF sputtering parameters to achieve a metatronics circuits based on ITO with opportunity tailored permittivity using the previously discussed process parameters. Therefore, we combine experimental data related to ENZ position with numerical approaches. For a homogeneous thickness of ITO with tailored values of real and imaginary part of the dielectric constant one can design optical lumped-circuit elements in an integrated system. For instance, using a similar approach proposed by Engheta et al in [43], we design and analyze a nanophotonic circuit based on our ITO films, whose permittivity is function of the oxygen flow-rate used for deposition. It is possible to use sub-wavelength photonic circuit based on different ITO film, sputtered at different oxygen flow-rate, for interacting with a propagating mode. A $\lambda$ = 1550 nm $TE_{10}$ mode is launched in a Silicon waveguide with a permittivity of 12. Our proof-of-principle metatronic circuit is comprised of two ITO-based nanostructures positioned in the center of the waveguide, in two possible configurations (parallel/series). The thickness, $t$, of the ITO films is 50 nm, thus a lumped circuit model can be applied, being $t \ll \lambda$. In the equivalent circuit, the film with real part of the permittivity ($\varepsilon'$) larger than zero acts as a capacitor, whereas the film with negative real part as an inductor. The films are also characterized by a damping ($\varepsilon'' > 0$), which induces losses, modeled in the lumped model as a resistances. A capacitor-like ITO film can be sputtered adopting a null oxygen flow-rate, obtaining an ENZ position shifted towards red with respect to the considered wavelength (1550 nm), while an inductor can be obtained using 20 sccm oxygen flow-rate, resulting into a blue shift of the ENZ position. When the films with $\varepsilon'$ with opposite signs are placed in parallel configuration, there is a pronounced impedance mismatch, leading to a high reflection coefficient, which translates to a -12 dB transmission of the signal. Contrarily, for films placed in series the imaginary part of the permittivity becomes negligible and only insertion losses are present, achieving a transmission of -4 dB. Based on our previous work on ITO-based electrooptic modulators [13, 53] our future work includes actively reconfigured metatronic circuit building blocks on-chip using our ITO film control.

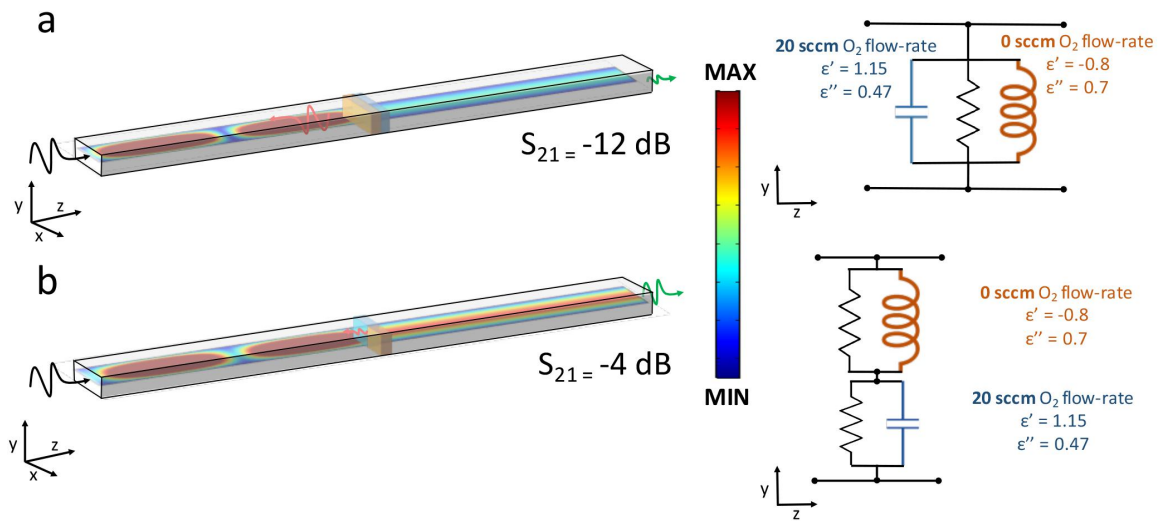

Figure 6. 3D view of the numerical simulation of a metatronics parallel and series setup. A $TE_{10}$ incident mode is propagating in a waveguide ($\varepsilon$ = 12). Two ITO film in parallel a and series b configuration are placed in between of the waveguide. Color map of the simulation results for the normalized electric field intensity distributions for the series (a) and parallel (b). The transmission coefficients ($S_{21}$) along with the equivalent circuit model and the process parameters used for obtaining specific permittivity values are reported on the right side.

**Discussion**

Understanding the effect of process parameters for tailoring ITO films electro-optical properties is crucial for both optimizing device performances such as in optics, electronics, or plasmonics, but also enables new material-platforms to realize metatronic circuits. In this study, we mapped the dependency of oxygen flow-rate and thermal annealing on the structural, optical and electrical properties of the ITO thin films deposited by RF sputtering. By tuning the oxygen flow rate we show control, after thermal treatment, of the absorption and the carrier concentration of the film. For thermally treated samples, we observed an overall increased absorption coefficient ($\kappa$), a 6-folds lower resistivity and significantly higher carrier concentration. Corroborated with modeling of the spectral response of the ITO films, we were able to precisely derive the resistivity and complex optical constant of annealed samples as function of the film depth, using a multi-layered fitting approach. For 10 sccm oxygen flow rate ITO film, we found the higher resistivity which was 10 times larger than the one obtained in null oxygen flow rate conditions, and the largest sweep throughout its depth. Ultimately, in annealed samples we found that the epsilon-near-zero reaches a maximum wavelength which red-shifts from 1.4 to 2.1 μm for 10 sccm oxygen flow-rate, only to blue-shift for higher oxygen concentration in the chamber, maintaining a similar trend with respect to the resistivity. We extend our study for different deposition time conditions aiming for mapping the ENZ position, damping and scattering time as function of oxygen flow-rate with deposition time. We simulate a sub-wavelength metatronic circuit based on ITO films, placed in the center of a waveguide, which exhibit tailored permittivity according to specific oxygen flow-rate. We demonstrated the possibility to use a lumped circuit entirely based on ITO as sub-wavelength filter at 1550 nm in an integrated photonic circuit.

In conclusion, in this novel study, we demonstrated the ability to accurately tune the electrical and optical properties of ITO films in a wide range of frequency ranging from VIS to IR. We thence show a precise control over conductivity, carrier density, and mobility of the ITO films enabled by tailoring of process parameters, such as oxygen flow rate, deposition time and post-deposition thermal treatment, assessed thorough cross-validation procedure. By controlling the oxygen flow ratio during sputtering, oxygen vacancies could be filled, making it possible to prepare optimal ITO films that exhibited high electrical performance, while still preserving optical transparency in a wide range of frequency. Following this approach, we demonstrate a fabricatability and repeatability of epsilon-near-zero (ENZ) on-chip platform. We anticipate these findings to enable a enables a plurality of functional devices for fields to include opto-electronic, plasmonic, metasurfaces, and possibly most novel, to metatronics. Using this rigorous material control, we finally give an example showing the ability to engineer a metatronic-based sub-wavelength building block demonstrating an optical switch.

**Methods**

**RF Deposition:** ITO ultra-thin films were deposited on a cleaned Si substrate with a nominal 300 nm SiO2 on it (1cm 1cm) at 313K by reactive RF sputtering using Denton Vacuum Discovery 550 Sputtering System. They were prepared with the same time which is 1500 seconds. The target is consisting of 10% SnO2 and 90% In2O3 by weight. The ITO films were prepared with the same Argon flow-rate which is 40 sccm and different oxygen flow-rates which are 0 sccm, 5 sccm, 10 sccm, 20 sccm, and 30 sccm respectively. All the deposition time is 1500s. There are four chips for each group. Two of them are tapped for later profilometer measurement. The vacuum setpoint is 5 Torr before deposition and the target will be pre-sputtered with the same deposition condition for 120s to remove the surface oxide layer of the target to avoid the contamination of the films. RF voltages were 300 voltage and RF bias were 25 voltage. After all parameters reached their set-points, deposition began.

**Annealing Process:** After deposition, few samples were annealed in a sealed chamber filled with $H_2$ and $N_2$ in order to avoid the influence of oxygen in the air during the annealing process at 350$^o$C for 15 min.

**Ellipsometry:** We carried out spectroscopic ellipsometry measurement using J.A. Woollam M-2000 DI, which covered wavelength from 200 nm to 1680 nm. Analysis of the data used the corresponding CompleteEASE to extract thickness, complex optical constants, and other electrical parameters. Silicon substrate and a nominal 300 nm silicon dioxide layer were also considered in ITO fitting model since, being commercial products, the related information were not exhaustive for analysis. We first fit the transparent region using Cauchy model to find out the closest thickness value for ITO thin film and fix it. Then we fitted the data using B-spline model and subsequently we expanded the fitting region from transparent region to the entire wavelength region. After that we re-parameterized the data using different oscillators in GenOsc, i.e. Drude, Cauchy-Lorentz and Lorentz oscillators. More details on the fitting model are provided in the SI.

**Resistance measurement** The resistivity and Hall voltages (Ecopia HMS-5500 Hall Effect measurement) of the films were measured using 4-terminal van der Pauw geometry. The electric contacts were ate corners of a 1x1 cm sample of the film to be investigated. Electrical measurements were used for defining the thin film sheet resistance, carrier concentration and mobility.

**Acknowledgements**

VS is funded by AFOSR (FA9550-17-1-0377) and ARO (W911NF-16-2-0194) and HD by NASA STTR, Phase I (80NSSC18P2146).